# Genuine tripartite entanglement in graviton-matter interactions

Pablo Guillermo Carmona Rufo[1,*], Anupam Mazumdar,[2] and Carlos Sabín[3]
[1]*Instituto de Física Teórica, UAM-CSIC, C/ Nicolás Cabrera 13-15, Campus de Cantoblanco, 28049 Madrid, Spain*
[2]*Van Swinderen Institute for Particle Physics and Gravity, University of Groningen, 9747 AG Groningen, The Netherlands*
[3]*Departamento de Física Teórica, Universidad Autónoma de Madrid, 28049 Madrid, Spain*

We demonstrate the existence of genuine tripartite non-Gaussian entanglement in a quantum gravitational system formed by a quantum harmonic oscillator coupled to a single frequency of a quantized gravitational wave. For this purpose, we introduce an entanglement witness, well suited for the graviton-matter interaction Hamiltonian analyzed here. We show that the witness is nonzero for the three-mode states generated by the Hamiltonian when the system starts in the ground state, thus proving the generation of genuine multipartite quantum entanglement at the most fundamental level of theoretical graviton-matter interactions.

## I. INTRODUCTION

The consistent implementation of the gravitational interaction into the quantum framework is considered to be one of the outstanding problems in current physics. Due to the lack of a direct experimental hint, the motivations for the need of such a theory were typically considered as purely theoretical. They suggest that the present model of physics could be fundamentally incomplete, but a direct experimental test of a full quantum gravitational theory would be out of reach.

In the last years, a novel approach to this open debate of quantum gravity has emerged, where the focus has shifted to just prove the quantum nature of gravity, without disclosing the underlying full quantum theory [1–7]. The rationale is that if entanglement can be generated solely by gravitational means, then gravity must be quantum. In this scheme, sometimes called QGEM (quantum gravity-induced entanglement of masses), the gravitational field is an effective linear quantum field which leaves its indirect trace in the dynamics of a pair of quantum systems, such as a couple of quantum harmonic oscillators. In previous works [8], QGEM is often studied by the introduction of an analogy to the field of optomechanics [9–12], since there are sometimes similarities between the Hamiltonian terms that arise in both areas.

Following this line of reasoning a step further, we notice that if two quantum systems get entangled by the exchange of a graviton this must mean that the graviton got entangled first with one of them, then transmitting the quantum correlations to the other—a familiar mechanism in the context of two-level atoms or qubits exchanging photons. Therefore, in this article, instead of tracing out the field and analyzing the dynamics of the quantum matter systems, we shift the focus to just a single matter system interacting with the quantum gravitational field. Due to the physical properties of the graviton field, we find that the most fundamental scenario involves three modes, namely the quantum oscillator mode and two graviton modes, each one representing a physical polarization. Moreover, the form of the graviton-matter interaction Hamiltonian gives rise to an interesting multipartite entanglement structure: The three-mode Hamiltonian consists of two-mode non-Gaussian interactions, as opposed to two-mode Gaussian [13,14] or direct three-mode interactions [15,16]. This suggests that previously known criteria based on Gaussian two-mode [17,18] or pure three-mode [19] correlations would fail to detect multipartite entanglement in this system.

We introduce an entanglement witness based on a suitable combination of two-mode third-order correlations in order to detect genuine tripartite entanglement in the family of states generated by the graviton-matter interaction Hamiltonian. We show that the witness is nonzero in a wide range of parameters when said Hamiltonian acts on an initial ground state, thus demonstrating the existence of genuine tripartite entanglement in this quantum gravitational system.

## II. INTERACTION HAMILTONIAN

We analyze the interaction between gravitons and quantum matter in the context of linearized quantum gravity. For this, we start by following Ref. [20], in which the authors consider a quantum harmonic oscillator coupled to quantized gravitational waves [21,22]. By the use of Fermi normal coordinates, the interaction Hamiltonian between graviton and matter degrees of freedom is found to be

$$H_{\text{int}} = \frac{m}{4} \frac{\partial h_{11}^2}{\partial t^2}\bigg|_{t=0} x^2, \quad (1)$$

where $m$ is the mass of the oscillator and $\ddot{h}_{11} = 2c^2 R_{0101}$ is the "+" component of the gravitational waves in the transverse traceless (TT) gauge, with $R_{0101}$ being the corresponding component of the Riemann tensor. Now, we can expand the gravitational field in plane waves [20,23,24],

$$\hat{h}_{ij} = \int d\mathbf{k} \sqrt{\frac{G\hbar}{\pi^2 c^2 \omega_k}} \hat{g}_{\mathbf{k},\lambda} e^\lambda_{ij}(\mathbf{n}) e^{-i(\omega_k t - \mathbf{k}\cdot\mathbf{x})} + \text{H.c.}, \quad (2)$$

*Contact author: pablog.carmona@estudiante.uam.es

where $\hat{g}_{k,\lambda}$ is the annihilation operator associated to the graviton field and $e^\lambda_{ij}(n)$ are the basis tensors for the two polarizations $\lambda = 1, 2$ (the summation over $\lambda$ is being implicitly assumed). Plugging this into (1), we get

$$H_{\text{int}} = \sum_\lambda \int d\mathbf{k}\, \mathcal{C}^\lambda_k \hat{g}_{k,\lambda} \hat{X}^2 + \text{H.c.}, \tag{3}$$

where we have introduced the adimensional amplitude quadrature

$$\hat{X} = \hat{b} + \hat{b}^\dagger, \tag{4}$$

which is related to the usual position observable through $\hat{x} = \delta_{\text{zpf}} \hat{X}$, where $\delta_{\text{zpf}} = \sqrt{\frac{\hbar}{2\mu\omega_m}}$, as well as

$$\mathcal{C}_\lambda = \sqrt{\frac{G\hbar^3 \omega_k^3}{64\pi^2 c^2 \omega_m^2}} e^\lambda_{11}(n). \tag{5}$$

Before we proceed any further, let us discuss an important point. While the work in Ref. [20] is presented in the continuous state space, our goal is to restrict ourselves to the discrete space by choosing just one graviton mode $\mathbf{k}$, drawing inspiration from the well-known single-mode models of quantum optics, such as Jaynes-Cummings or Rabi Hamiltonians. In the continuum regime, that is, the limit $V \to \infty$, the sum over $\mathbf{k}$ modes is replaced by an integral [25],

$$\frac{1}{V} \sum_{\mathbf{k}} \xrightarrow[V \to \infty]{} \int \frac{d\mathbf{k}}{(2\pi)^3}. \tag{6}$$

Therefore, we need to substitute said integral by the right-hand side of (6), taking into account that, while the $\hat{g}_{k,\lambda}$ operators are adimensional in the discrete space, they have units of $\sqrt{V} = L^{3/2}$ in the continuum, which means that they must be made dimensionless by means of $\hat{g}_{k,\lambda} \to \frac{\sqrt{(2\pi)^3}}{\sqrt{V}} \hat{g}_{k,\lambda}$, and therefore $\mathcal{C}_\lambda \to \frac{\sqrt{(2\pi)^3}}{\sqrt{V}} \mathcal{C}_\lambda$. We are choosing the wave vector for the fundamental mode of a system confined in a cube of side $L$, so we will have $k = |\vec{k}| = \frac{2\pi}{L} \to \sqrt{V} = (\frac{2\pi c}{\omega_k})^{3/2}$ and thus

$$\mathcal{C}'_\lambda = \sqrt{\frac{G\hbar^3 \omega_k^6}{64\pi^2 c^5 \omega_m^2}} e^\lambda_{11}(n). \tag{7}$$

Therefore, we end up expressing our interaction Hamiltonian as

$$H_{\text{int}} = [\mathcal{C}'_1(\hat{g}_1 + \hat{g}_1^\dagger) + \mathcal{C}'_2(\hat{g}_2 + \hat{g}_2^\dagger)] \hat{X}^2. \tag{8}$$

Grouping the terms in a suitable way, we get

$$\begin{aligned} H_{\text{int}} = \mathcal{C}'_1 &[(\hat{g}_1 \hat{b}^2 + \hat{g}_1^\dagger \hat{b}^{\dagger 2}) + (\hat{g}_1 \hat{b}^{\dagger 2} + \hat{g}_1^\dagger \hat{b}^2). \\ &+ (\hat{g}_1 \hat{b} \hat{b}^\dagger + \hat{g}_1^\dagger \hat{b}^\dagger \hat{b}) + (\hat{g}_1 \hat{b}^\dagger \hat{b} + \hat{g}_1^\dagger \hat{b} \hat{b}^\dagger)] \\ + \mathcal{C}'_2 &[(\hat{g}_2 \hat{b}^2 + \hat{g}_2^\dagger \hat{b}^{\dagger 2}) + (\hat{g}_2 \hat{b}^{\dagger 2} + \hat{g}_2^\dagger \hat{b}^2). \\ &+ (\hat{g}_2 \hat{b} \hat{b}^\dagger + \hat{g}_2^\dagger \hat{b}^\dagger \hat{b}) + (\hat{g}_2 \hat{b}^\dagger \hat{b} + \hat{g}_2^\dagger \hat{b} \hat{b}^\dagger)]. \end{aligned} \tag{9}$$

This is a three-mode Hamiltonian consisting of combinations of two-mode interactions which are linear in the graviton modes and quadratic in the oscillator mode. In order to detect tripartite entanglement [17,18] in three-mode quantum systems, the conventional criteria are based on inequalities depending on expectation values of certain operators and correlations that involve the three modes pair by pair, such as $\langle x_i x_j \rangle$ [18]. This is convenient for Gaussian states, which can be fully described by the covariance matrix elements. Nonetheless, it has been already showed in Refs. [15,16] that some of these inequalities are not fitted to spot the tripartite entanglement generated by every single three-mode Hamiltonian. In Ref. [26], this is put into practice by showing that this criterion is not effective when trying to find tripartite entanglement in the states generated by the three-mode spontaneous parametric down-conversion (SPDC) Hamiltonian given by $H = \frac{\hbar g_0}{2}(abc + a^\dagger b^\dagger c^\dagger)$. The reason is that the Hamiltonian consists of pure three-mode interactions, not combinations of pairwise ones and therefore the entanglement witness must have a similar structure.

Before proving the presence of entanglement in our system, we show the form of the family of states generated by the Hamiltonian in (9) in a perturbative regime. The time evolution of the initial state will be given by

$$|\psi(t)\rangle = e^{-iH_{\text{int}} t/\hbar} |\psi_0\rangle. \tag{10}$$

We need to choose any nonentangled state as our initial state, since we are interested in showing the generation of entanglement by proving that the Hamiltonian is able to entangle it. The vacuum state is the most natural choice, so we assume

$$|\psi_0\rangle = |0\rangle_{\hat{g}_1} |0\rangle_{\hat{g}_2} |0\rangle_m. \tag{11}$$

We express our Hamiltonian as

$$H_{\text{int}} = [\mathcal{C}'_1(\hat{g}_1 + \hat{g}_1^\dagger) + \mathcal{C}'_2(\hat{g}_2 + \hat{g}_2^\dagger)] \hat{X}^2 \equiv \mathcal{C}'_1 H_1 + \mathcal{C}'_2 H_2, \tag{12}$$

which allows us to write the time evolution operator as

$$e^{-i(\varepsilon_1 H_1 + \varepsilon_2 H_2)}, \tag{13}$$

where $\varepsilon_i = \mathcal{C}'_i t/\hbar$. We now perform a perturbative expansion on the time evolution operator around $\varepsilon_i$ in order to find the time-evolved state,

$$e^{-iHt/\hbar} \simeq \mathbb{1} - i(\varepsilon_1 H_1 + \varepsilon_2 H_2) + \mathcal{O}(\varepsilon^2). \tag{14}$$

We find that the only terms that do not cancel when acting upon the ground state are

$$\begin{aligned} |\psi\rangle &\simeq [\mathbb{1} - i(\varepsilon_1 H_1 + \varepsilon_2 H_2)] |\psi_0\rangle \\ &= |000\rangle - i[\varepsilon_1(g_1^\dagger \hat{b}^{\dagger 2} + g_1^\dagger \hat{b}\hat{b}^\dagger) \\ &\quad + \varepsilon_2(g_2^\dagger \hat{b}^{\dagger 2} + g_2^\dagger \hat{b}\hat{b}^\dagger)] |000\rangle \\ &= |000\rangle - i[\varepsilon_1(|100\rangle + \sqrt{2}|102\rangle) \\ &\quad + \varepsilon_2(|010\rangle + \sqrt{2}|012\rangle)]. \end{aligned} \tag{15}$$

### III. ENTANGLEMENT WITNESS

In order to achieve the detection of entanglement in this family of states, we follow a strategy similar to the one in Ref. [26], trying to find a suitable entanglement witness with an structure reflecting the properties of (9), which contains only two-mode interactions but beyond the Gaussian formalism. We start by studying the correlations between all possible bipartitions of the system and considering the inequalities shown in Ref. [27] for each of them. These inequalities state that if we consider the total Hilbert space of our system as $\mathcal{H} = \mathcal{H}_1 \otimes \mathcal{H}_2$, where the 1 and 2 indices denote its two subsystems, and we let $A^{(1)}$ be an operator acting on $\mathcal{H}_1$ and $A^{(2)}$ an operator on $\mathcal{H}_2$, then the total state of the system will not be entangled with respect to this partition if

$$|\langle A^{(1)} A^{(2)} \rangle| \leqslant \sqrt{\langle A^{(1)\dagger} A^{(1)} \rangle \langle A^{(2)\dagger} A^{(2)} \rangle}. \qquad (16)$$

The condition expressed in (16) means that we can define $I = |\langle A^{(1)} A^{(2)} \rangle| - \sqrt{\langle A^{(1)\dagger} A^{(1)} \rangle \langle A^{(2)\dagger} A^{(2)} \rangle}$, which will give us the condition that the state is fully inseparable if $I > 0$ for the three possible bipartitions of our tripartite system. In other words, if we are able to find a way to divide our three-mode Hilbert space into two subspaces, as well as find operators $A^{(1)}$ and $A^{(2)}$ acting on them that make $I > 0$ for the three possible bipartitions, we will have proven the full inseparability of the system. We note that, in order to calculate the expected value of any observable $\mathcal{O}$, if the system starts in the ground state $|0\rangle$, due to the weakness of the gravitational interaction, we can use perturbation theory to write [16]

$$\langle \mathcal{O} \rangle = \langle 0|\mathcal{O}|0\rangle + i\frac{t}{\hbar}\langle 0|[H_{\text{int}}, \mathcal{O}]|0\rangle + O(\lambda^2), \qquad (17)$$

where $\lambda$ is a perturbative parameter. We find that, choosing $A^{(1)}$ and $A^{(2)}$ in such a way that $A^{(1)} A^{(2)} = (\mathbb{1} + \hat{g}_1)(\mathbb{1} + \hat{g}_2)\hat{b}^2$ for the three bipartitions, the left-hand side of (16) will read, using (17),

$$|\langle(\mathbb{1} + \hat{g}_1)(\mathbb{1} + \hat{g}_2)\hat{b}^2\rangle| = 2|\mathcal{C}'_1 + \mathcal{C}'_2|\frac{t}{\hbar} \geqslant 0, \qquad (18)$$

while the right-hand side will vanish, violating the given condition and therefore assuring the full inseparability of the state (see Appendix A for more details). Nevertheless, even if a state generated by our graviton-matter Hamiltonian has full inseparability, there is still the possibility that it can be decomposed in the following way,

$$\rho = P_1 \rho_1^{(a)} \otimes \rho_1^{(bc)} + P_2 \rho_1^{(b)} \otimes \rho_1^{(ac)} + P_3 \rho_1^{(c)} \otimes \rho_1^{(ab)}, \qquad (19)$$

with $P_1 + P_2 + P_3 = 1$. If that is the case, we would not classify the inseparability of the state as a form of tripartite entanglement. Therefore, we are interested in studying genuine tripartite entanglement, which is defined as the property of fully inseparable states that cannot be written as in (19) [18,28,29]. Thus, we would like to find a condition that allows us to determine whether our interaction Hamiltonian produces genuine tripartite entanglement. For that purpose, we can begin by making use of the triangle inequality to write

$$|\langle A^{(1)} A^{(2)} \rangle_\rho| \leqslant P_1 |\langle A^{(1)} A^{(2)} \rangle_{\rho_1}| + P_2 |\langle A^{(1)} A^{(2)} \rangle_{\rho_2}| + P_3 |\langle A^{(1)} A^{(2)} \rangle_{\rho_3}|, \qquad (20)$$

where we are defining

$$\rho_1 = \rho_1^{(a)} \otimes \rho_1^{(bc)}, \quad \rho_2 = \rho_2^{(b)} \otimes \rho_2^{(ac)}, \quad \rho_3 = \rho_3^{(c)} \otimes \rho_3^{(ab)}. \qquad (21)$$

We also know that, by construction, $\rho_1$, $\rho_2$, and $\rho_3$ are biseparable and therefore must follow the inequality in (16). Thus,

the inequality in our case would look like

$$|\langle(\mathbb{1} + \hat{g}_1)(\mathbb{1} + \hat{g}_2)\hat{b}^2\rangle|$$
$$\leqslant P_1 \sqrt{\langle(\mathbb{1} + \hat{g}_1^\dagger)(\mathbb{1} + \hat{g}_1)\rangle_{\rho_1} \langle(\mathbb{1} + \hat{g}_2^\dagger)(\mathbb{1} + \hat{g}_2)\hat{b}^{\dagger 2}\hat{b}^2\rangle_{\rho_1}}$$
$$+ P_2 \sqrt{\langle(\mathbb{1} + \hat{g}_2^\dagger)(\mathbb{1} + \hat{g}_2)\rangle_{\rho_2} \langle(\mathbb{1} + \hat{g}_1^\dagger)(\mathbb{1} + \hat{g}_1)\hat{b}^{\dagger 2}\hat{b}^2\rangle_{\rho_2}}$$
$$+ P_3 \sqrt{\langle(\mathbb{1} + \hat{g}_1^\dagger)(\mathbb{1} + \hat{g}_1)(\mathbb{1} + \hat{g}_2^\dagger)(\mathbb{1} + \hat{g}_2)\rangle_{\rho_3} \langle \hat{b}^{\dagger 2}\hat{b}^2 \rangle_{\rho_3}}. \qquad (22)$$

Using (19), we can write that, for example,

$$P_1 \langle(\mathbb{1} + \hat{g}_1^\dagger)(\mathbb{1} + \hat{g}_1)\rangle_{\rho_1}$$
$$= \langle(\mathbb{1} + \hat{g}_1^\dagger)(\mathbb{1} + \hat{g}_1)\rangle_\rho - P_2 \langle(\mathbb{1} + \hat{g}_1^\dagger)(\mathbb{1} + \hat{g}_1)\rangle_{\rho_2}$$
$$- P_3 \langle(\mathbb{1} + \hat{g}_1^\dagger)(\mathbb{1} + \hat{g}_1)\rangle_{\rho_3} \leqslant \langle(\mathbb{1} + \hat{g}_1^\dagger)(\mathbb{1} + \hat{g}_1)\rangle_\rho, \qquad (23)$$

and equivalently with the other two expected values. This allows us to claim that if a state produced by our Hamiltonian can be written in the form of (19) and thus does not possess genuine tripartite entanglement, it will fulfill

$$|\langle(\mathbb{1} + \hat{g}_1)(\mathbb{1} + \hat{g}_2)\hat{b}^2\rangle|$$
$$\leqslant \sqrt{\langle(\mathbb{1} + \hat{g}_1^\dagger)(\mathbb{1} + \hat{g}_1)\rangle \langle(\mathbb{1} + \hat{g}_2^\dagger)(\mathbb{1} + \hat{g}_2)\hat{b}^{\dagger 2}\hat{b}^2\rangle}$$
$$+ \sqrt{\langle(\mathbb{1} + \hat{g}_2^\dagger)(\mathbb{1} + \hat{g}_2)\rangle \langle(\mathbb{1} + \hat{g}_1^\dagger)(\mathbb{1} + \hat{g}_1)\hat{b}^{\dagger 2}\hat{b}^2\rangle}$$
$$+ \sqrt{\langle(\mathbb{1} + \hat{g}_1^\dagger)(\mathbb{1} + \hat{g}_1)(\mathbb{1} + \hat{g}_2^\dagger)(\mathbb{1} + \hat{g}_2)\rangle \langle \hat{b}^{\dagger 2}\hat{b}^2 \rangle}, \qquad (24)$$

which allows us to define

$$\mathcal{G}_1 = |\langle(\mathbb{1} + \hat{g}_1)(\mathbb{1} + \hat{g}_2)\hat{b}^2\rangle| - \mathcal{O}_1 - \mathcal{O}_2 - \mathcal{O}_3, \qquad (25)$$

where the $\mathcal{O}_i$ are each of the square root terms in (24). In principle, we could already use this witness in order to detect the entanglement in our system. However, following the reasoning in Ref. [19], we can also exploit the fact that the expectation values of a mixed state cannot be larger than the largest of its components, in order to find the following improved witness,

$$\mathcal{G}_2 = |\langle(\mathbb{1} + \hat{g}_1)(\mathbb{1} + \hat{g}_2)\hat{b}^2\rangle| - \max(\mathcal{O}_1, \mathcal{O}_2, \mathcal{O}_3). \qquad (26)$$

Computing all the relevant terms, we conclude that all the square root terms vanish with these choices for $A^{(1)}$ and $A^{(2)}$ (see Appendix A). Due to this, either witness could be used indifferently in the context of this work. Thus, we conclude that our witnesses, when evaluated over the family of states generated by the action of the Hamiltonian (9) over an initial vacuum state, are equal to a positive semidefinite quantity $\mathcal{G}$,

$$\mathcal{G}_1 = \mathcal{G}_2 \equiv \mathcal{G} = |\langle(\mathbb{1} + \hat{g}_1)(\mathbb{1} + \hat{g}_2)\hat{b}^2\rangle| = \Omega \cdot t \geqslant 0, \qquad (27)$$

with

$$\Omega = \frac{2|\mathcal{C}'_1 + \mathcal{C}'_2|}{\hbar} = \sqrt{\frac{G\hbar \omega_k^6}{16\pi^2 c^5 \omega_m^2}} |e_{11}^1 + e_{11}^2|. \qquad (28)$$

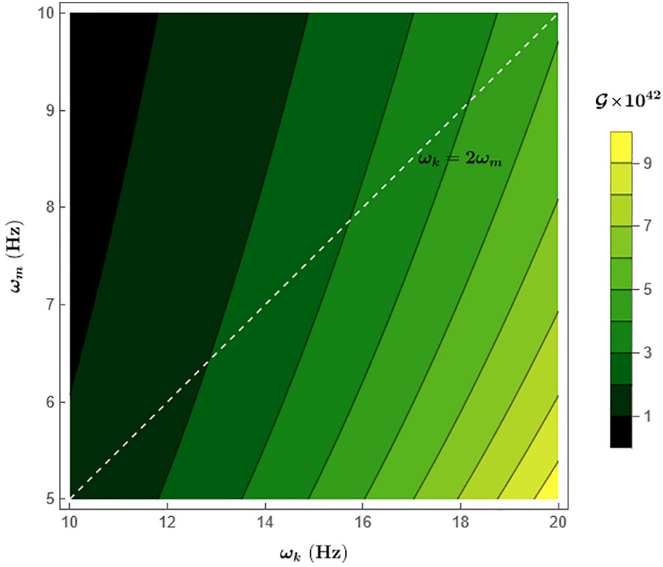

FIG. 1. Value of the genuine tripartite entanglement witness $\mathcal{G}$ as a function of $\omega_k$ and $\omega_m$ for $t = 1$ s and $e_{11}^1 = e_{11}^2 = 1/\sqrt{2}$.

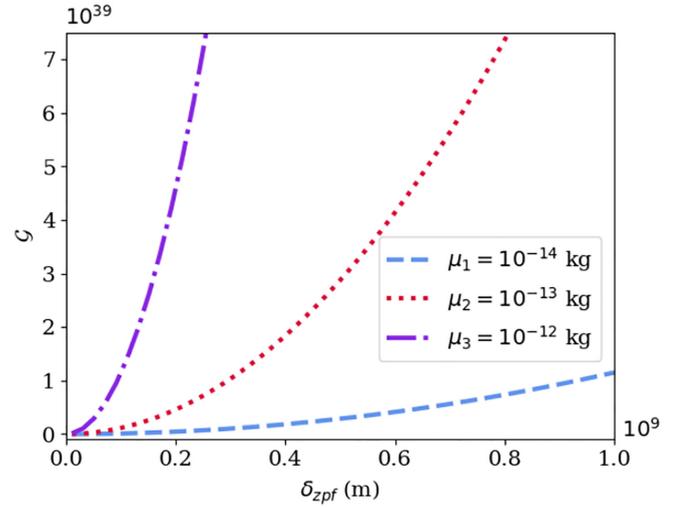

FIG. 2. Value of the genuine tripartite entanglement witness $\mathcal{G}$ as a function of the zero-point fluctuation of the matter system $\delta_{\text{zpf}}$, for three different values of the mass $\mu$ and $\omega_k = 10$ Hz.

## IV. RESULTS

In Fig. 1, we present the result for the entanglement witness for different values of $\omega_k$ and $\omega_m$ and $e_{11}^1 = e_{11}^2 = 1/\sqrt{2}$ (see Appendix B for a discussion on the polarization tensor). The intervals for the frequencies are in accordance to what is done in Ref. [20], where the authors estimate the order of magnitude of gravitational decoherence in the QGEM protocol [1], for which they consider an effective harmonic trap frequency of $\omega_m \sim 2\pi$ Hz. We choose a similar range for $\omega_k$ as well, considering that the Hamiltonian (9) includes terms such as $\hat{g}_1 \hat{b}^{\dagger 2} + \hat{g}_1^{\dagger} \hat{b}^2$, etc., that would be energy conserving for $\omega_k = 2\omega_m$. These terms would be favored in the evolution eventually, and indeed would be the only ones relevant in a master-equation Born-Markov approach [20]—a scenario resembling the familiar rotating-wave approximation in quantum optics. As shown, the entanglement witness reaches very small values in this parameter range, of the order of $10^{-42}$. However, obviously other parameter ranges might be considered as well, corresponding to different parts of the gravitational wave spectrum and different models of quantum harmonic oscillator. Notice that with the condition $\omega_k = 2\omega_m$, the entanglement witness would be proportional to $\omega_k^2 t$.

In Fig. 2, we show the dependence of the entanglement witness $\mathcal{G}$ with respect to the zero-point fluctuation energy $\delta_{\text{zpf}}$ of the system for three different values of the reduced mass $\mu$ of the harmonic oscillator. The parameter values are chosen following a similar reasoning as for the case of Fig. 1: The mass values are similar to the one considered in Ref. [20] in order to estimate gravitational decoherence in the QGEM protocol, while the range selected for $\delta_{\text{zpf}}$ allows $\omega_m$ to stay between 0 and $10^2$ Hz for those values of $\mu$.

As it can be sensed in the graph, the witness scales as $\delta_{\text{zpf}}^2$, since from (27) and (28) we notice that $\mathcal{G} \sim \omega_m^{-1}$, while $\delta_{\text{zpf}} \sim \omega_m^{-2}$. This result is of particular interest since it reminds us of the behavior of entanglement entropy, which is known to be linear with the area in a range of quantum physical systems [30,31]. However, it is important not to assume a direct relationship between both magnitudes, as the witness does not necessarily grow with the amount of entanglement, it just signals its presence in the case that the obtained result is positive.

## V. CONCLUSIONS

Summarizing, we show that the interaction Hamiltonian corresponding to the system composed of a quantum harmonic oscillator and a gravitational field with two polarizations is able to generate states with full inseparability and genuine tripartite entanglement. The class of entanglement that shows up in these states is different to the one displayed in other tripartite setups, such as Ref. [26] (where the interaction involves the three modes at the same time) or Ref. [14] (where the interactions are two-mode and Gaussian) and thus we introduce an entanglement witness, composed of a combination of non-Gaussian pairwise correlations, reflecting the structure of the corresponding three-mode graviton-matter interaction Hamiltonian.

In addition to the theoretical interest of the generation of graviton matter at a fundamental level, we notice that this entanglement is actually the hidden resource for the generation of entanglement in QGEM setups, since the graviton-oscillator entanglement can later be transferred to another oscillator. We also believe that these results might prepare the ground for the appropriate experimental entanglement tests, similarly to the work in Ref. [16]. Despite the fact that the detection of a graviton field in the laboratory is still a long way ahead of us, an experimental simulation that recreates the properties of the system could be of interest. Indeed, in Ref. [16], besides three-mode SPDC, a two-mode Hamiltonian with a structure similar to a two-mode sector of (9) was implemented, suggesting a clear path for the experimental realization of an interaction analog to the full Hamiltonian (9). Moreover, the digitization of the system in order to launch a digital quantum simulation in a quantum computer, following the spirit of Refs. [8,32] for QGEM setups, seems also a feasible research path to explore in the near future.

## ACKNOWLEDGMENTS

P.G.C.R. acknowledges Grant No. PRE2022-102488 funded by MCIN/AEI/10.13039/501100011033 and FSE+, with Project Code No. PID2021-127726NB-I00. C.S. acknowledges financial support through the Ramón y Cajal Programme (RYC2019-028014-I). A.M.'s research is partly funded by the Gordon and Betty Moore Foundation through Grant No. GBMF12328 [Grant DOI: 10.37807/GBMF12328].

## APPENDIX A: CALCULATION OF THE ENTANGLEMENT WITNESS

In this Appendix, we will provide the detailed steps of some of the calculations in the main body of the paper. The left-hand side of the entanglement witness defined in (25) is calculated as

$$\langle(\mathbb{1}+\hat{g}_1)(\mathbb{1}+\hat{g}_2)\hat{b}^2\rangle$$
$$= \cancel{\langle 0|(\mathbb{1}+\hat{g}_1)(\mathbb{1}+\hat{g}_2)\hat{b}^2|0\rangle}$$
$$- i\frac{t}{\hbar}\langle 0|(\mathbb{1}+\hat{g}_1)(\mathbb{1}+\hat{g}_2)\hat{b}^2 H_{\text{int}}|0\rangle$$
$$+ i\frac{t}{\hbar}\cancel{\langle 0|H_{\text{int}}(\mathbb{1}+\hat{g}_1)(\mathbb{1}+\hat{g}_2)\hat{b}^2|0\rangle}$$
$$= -i\frac{t}{\hbar}\langle 0|(\hat{g}_1+\hat{g}_2)\hat{b}^2(\mathcal{C}'_1\hat{g}^\dagger_1\hat{b}^{\dagger 2}+\mathcal{C}'_2\hat{g}^\dagger_2\hat{b}^{\dagger 2})|0\rangle$$
$$= -i\frac{t}{\hbar}(\mathcal{C}'_1\langle 0|\hat{g}_1\hat{g}^\dagger_1\hat{b}^2\hat{b}^{\dagger 2}|0\rangle + \mathcal{C}'_2\langle 0|\hat{g}_2\hat{g}^\dagger_2\hat{b}^2\hat{b}^{\dagger 2}|0\rangle)$$
$$= -2i\frac{t}{\hbar}(\mathcal{C}'_1+\mathcal{C}'_2). \tag{A1}$$

Thus:

$$|\langle(\mathbb{1}+\hat{g}_1)(\mathbb{1}+\hat{g}_2)\hat{b}^2\rangle| = 2|\mathcal{C}'_1+\mathcal{C}'_2|\frac{t}{\hbar}. \tag{A2}$$

On the other hand, it was mentioned that all of the square root terms in our expression for the witness, which are given in (24), always become zero for the family of states produced by our Hamiltonian, since

$$\langle(\mathbb{1}+\hat{g}^\dagger_2)(\mathbb{1}+\hat{g}_2)\hat{b}^{\dagger 2}\hat{b}^2\rangle = 0$$
$$\Rightarrow \sqrt{\langle(\mathbb{1}+\hat{g}^\dagger_1)(\mathbb{1}+\hat{g}_1)\rangle\langle(\mathbb{1}+\hat{g}^\dagger_2)(\mathbb{1}+\hat{g}_2)\hat{b}^{\dagger 2}\hat{b}^2\rangle} = 0,$$
$$\langle(\mathbb{1}+\hat{g}^\dagger_1)(\mathbb{1}+\hat{g}_1)\hat{b}^{\dagger 2}\hat{b}^2\rangle = 0$$
$$\Rightarrow \sqrt{\langle(\mathbb{1}+\hat{g}^\dagger_2)(\mathbb{1}+\hat{g}_2)\rangle\langle(\mathbb{1}+\hat{g}^\dagger_1)(\mathbb{1}+\hat{g}_1)\hat{b}^{\dagger 2}\hat{b}^2\rangle} = 0,$$
$$\langle\hat{b}^{\dagger 2}\hat{b}^2\rangle = 0$$
$$\Rightarrow \sqrt{\langle(\mathbb{1}+\hat{g}^\dagger_1)(\mathbb{1}+\hat{g}_1)(\mathbb{1}+\hat{g}^\dagger_2)(\mathbb{1}+\hat{g}_2)\rangle\langle\hat{b}^{\dagger 2}\hat{b}^2\rangle} = 0.$$
$$\tag{A3}$$

## APPENDIX B: POLARIZATION TENSOR

The polarization tensor $e_{\mu\nu}$ for the graviton field arises from the plane-wave solutions to the field equations of the gravitational field $h_{\mu\nu}$. Out of all its components, only two represent physically significant degrees of freedom, which are typically chosen to be $e_{11}$ and $e_{12}$ [24]. We must also take into account the completeness relation for these tensor components given in Ref. [20],

$$\sum_\lambda e^\lambda_{ij} e^\lambda_{kl} = P_{ik}P_{jl} + P_{il}P_{jk} - P_{ij}P_{kl}, \tag{B1}$$

where $P_{ij} = P_{ij}(\mathbf{n}) = \delta_{ij} - \mathbf{n}_i\mathbf{n}_j$. From (1) and (2) we see that only $e^\lambda_{11}$ will be relevant for our system, which means that we can write

$$\left(e^1_{11}\right)^2 + \left(e^2_{11}\right)^2 = P_{11}P_{11}. \tag{B2}$$

In Ref. [20], they give a result for the value of this quantity averaged over all possible directions of $\mathbf{k}$. However, as we have already discussed, we are working on just one graviton mode. By choosing the direction of that mode to be $\mathbf{n} = \mathbf{u}_3$, we get $P_{11}P_{11} = 1$ and thus

$$\left(e^1_{11}\right)^2 + \left(e^2_{11}\right)^2 = 1. \tag{B3}$$

In order to perform the numerical simulation carried out in Fig. 1, we needed to choose a value for these two magnitudes. We decided to pick the values $e^1_{11} = e^2_{11} = 1/\sqrt{2}$, which fulfill the necessary conditions for the polarization tensor and give us the quantity $|e^1_{11} + e^2_{11}| = \sqrt{2}$, which we plugged in our result for the entanglement witness to perform the necessary calculations.

[1] S. Bose, A. Mazumdar, G. W. Morley, H. Ulbricht, M. Toroš, M. Paternostro, A. A. Geraci, P. F. Barker, M. Kim, and G. Milburn, Phys. Rev. Lett. **119**, 240401 (2017).

[2] C. Marletto and V. Vedral, Phys. Rev. Lett. **119**, 240402 (2017).

[3] S. Bose, A. Mazumdar, M. Schut, and M. Toroš, Phys. Rev. D **105**, 106028 (2022).

[4] D. Biswas, S. Bose, A. Mazumdar, and M. Toroš, Phys. Rev. D **108**, 064023 (2023).

[5] D. Carney, P. C. Stamp, and J. M. Taylor, Class. Quantum Grav. **36**, 034001 (2019).

[6] D. Carney, Phys. Rev. D **105**, 024029 (2022).

[7] R. J. Marshman, A. Mazumdar, and S. Bose, Phys. Rev. A **101**, 052110 (2020).

[8] P. G. C. Rufo, A. Mazumdar, S. Bose, and C. Sabín, EPJ Quantum Technol. **11**, 31 (2024)

[9] S. Bose, K. Jacobs, and P. L. Knight, Phys. Rev. A **56**, 4175 (1997).

[10] M. Aspelmeyer, T. J. Kippenberg, and F. Marquardt, Rev. Mod. Phys. **86**, 1391 (2014).

[11] C. K. Law, Phys. Rev. A **51**, 2537 (1995).

[12] A. Ferreri, H. Pfeifer, F. K. Wilhelm, S. Hofferberth, and D. E. Bruschi, Phys. Rev. A **106**, 033502 (2022).

[13] S. Gerke, J. Sperling, W. Vogel, Y. Cai, J. Roslund, N. Treps, and C. Fabre, Phys. Rev. Lett. **114**, 050501 (2015).

[14] C. W. S. Chang, M. Simoen, J. Aumentado, C. Sabín, P. Forn-Díaz, A. M. Vadiraj, F. Quijandría, G. Johansson, I. Fuentes, and C. M. Wilson, Phys. Rev. Appl. **10**, 044019 (2018).

[15] E. A. R. González, A. Borne, B. Boulanger, J. Levenson, and K. Bencheikh, Phys. Rev. Lett. **120**, 043601 (2018).

[16] C. W. S. Chang, C. Sabín, P. Forn-Díaz, F. Quijandría, A. M. Vadiraj, I. Nsanzineza, G. Johansson, and C. M. Wilson, Phys. Rev. X **10**, 011011 (2020).

[17] P. van Loock and A. Furusawa, Phys. Rev. A **67**, 052315 (2003).

[18] R. Y. Teh and M. D. Reid, Phys. Rev. A **90**, 062337 (2014).

[19] A. A. Casado and C. Sabín, Phys. Rev. A **105**, 022401 (2022).

[20] M. Toroš, A. Mazumdar, and S. Bose, Phys. Rev. D **109**, 084050 (2024).

[21] S. Bose, A. Mazumdar, and M. Toroš, Phys. Rev. D **104**, 066019 (2021).

[22] S. Bose, A. Mazumdar, and M. Toroš, Nucl. Phys. B **977**, 115730 (2022).

[23] T. Oniga and C. H.-T. Wang, Phys. Rev. D **93**, 044027 (2016).

[24] S. Weinberg, *Gravitation and Cosmology: Principles and Applications of the General Theory of Relativity* (Wiley, New York, 1972).

[25] D. Craig and T. Thirunamachandran, *Molecular Quantum Electrodynamics: An Introduction to Radiation-molecule Interactions*, Dover Books on Chemistry Series (Dover, New York, 1998).

[26] A. Agustí, C. W. Chang, F. Quijandría, G. Johansson, C. M. Wilson, and C. Sabín, Phys. Rev. Lett. **125**, 020502 (2020).

[27] H. Zheng, H. T. Dung, and M. Hillery, Phys. Rev. A **81**, 062311 (2010).

[28] S. Gerke, J. Sperling, W. Vogel, Y. Cai, J. Roslund, N. Treps, and C. Fabre, Phys. Rev. Lett. **117**, 110502 (2016).

[29] E. Shchukin and P. van Loock, Phys. Rev. A **92**, 042328 (2015).

[30] J. Eisert, M. Cramer, and M. B. Plenio, Rev. Mod. Phys. **82**, 277 (2010).

[31] Y. Ge and J. Eisert, New J. Phys. **18**, 083026 (2016).

[32] C. Sabín, EPJ Quantum Technol. **10**, 4 (2023).